 \documentclass[amsmath,amssymb,showkeys,aps,unsortedaddress,prb,twocolumn,preprintnumbers]{revtex4}

\usepackage{graphicx}
\usepackage{dcolumn}
\usepackage{bm}
\usepackage[usenames,dvipsnames]{color}

\def\tens#1{\ensuremath{\mathsf{#1}}}
\newcommand{\J}[0]{\ensuremath{\mathfrak J}}
\newlength{\largFigSeb} 

\def\scpDIAS{%
School of Cosmic Physics, Dublin Institute for Advanced Studies,
Ireland}
\def\luthOBS{%
Laboratoire de l'Univers et ses Th\'eories, Observatoire de Paris,
France}
\def\dptaCEA{%
D\'epartement de Physique Th\'eorique et Appliqu\'ee, Commissariat
\`a l'\'Energie Atomique, Bruy\`eres-Le-Ch\^atel, France}
\def\daUMA{%
Department of Astronomy, University of Maryland, USA}

\begin{document}
\preprint{Physics of Plasmas, {\bf 13}, 113301 (2006).}

\title{Modeling multidimensional effects in the propagation of
radiative shocks}

\author{S\'ebastien Leygnac} \email{sleygnac@cp.dias.ie}
\affiliation{\scpDIAS}
 \affiliation{\luthOBS}
\author{Laurent Boireau}
\affiliation{\luthOBS}
\affiliation{\dptaCEA}
\author{Claire Michaut}
\affiliation{\luthOBS}
\author{Thierry Lanz}
\affiliation{\daUMA}
\affiliation{\luthOBS}
\author{Chantal Stehl\'e}
\affiliation{\luthOBS}
\author{Christine Clique}
\author{Serge Bouquet}
\affiliation{\dptaCEA}

\date{\today}

\begin{abstract}
Radiative shocks (also called supercritical shocks) are high Mach
number shock waves that photoionize the medium ahead of the shock
front and give rise to a radiative precursor. They are generated in
the laboratory using high-energy or high-power lasers and are
frequently present in a wide range of astronomical objects.
Their modelisation in one dimension has been the subject of numerous
studies, but generalization to three dimensions is not
straightforward.
We calculate analyticaly the absorption of radiation in a grey
uniform cylinder and show how it decreases with $\chi R$, the
product of the opacity $\chi$ and of the cylinder radius $R$. Simple
formulas, whose validity range increases when $\chi R$ diminishes,
are derived for the radiation field on the axis of symmetry.
Numerical calculations in three dimensions of the radiative energy
density, flux and pressure created by a stationary shock wave show
how the radiation decreases whith $R$.
Finally, the bidimensional structures of both the precursor and the
radiation field are calculated with time-dependent radiation
hydrodynamics numerical simulations and the influence of
two-dimensional effects on the electron density, the temperature,
the shock velocity and the shock geometry are exhibited.
These simulations show how the radiative precursor shortens, cools
and slows down when $R$ is decreased.
\end{abstract}

\maketitle

%
%
\section{\label{sec:Intro}Introduction}

In many astrophysical systems, the effects of radiation on
hydrodynamics are strong.
This is the case with fast outflows and shocks, such as those
encountered in jets, bow shocks produced by the interaction of
jets with the surrounding interstellar
medium,\cite{2002AJ....123..362R}
or radiative shocks arising in the envelope of pulsating evolved
stars.\cite{1997A&A...324.1046H}
Nowadays, these processes can be observed with an increasing level
of details. For example, high angular resolution imaging of jets
produced by Young Stellar Objects \cite{2002AJ....123..362R} shows
complex structures of bright knots and shocks inside the jets.
The improvement of observation techniques, and more especially the
combination of spectroscopy and high angular resolution techniques,
will allow the study of the physical properties of shocks, in addition
to the study of their complex shapes. This combination will be
achieved by AMBER \cite{2004SPIE.5491.1742G} on the {\it Very
Large Telescope Interferometer} for
instance.

However, the interpretation of these new data should go together
with the improvement and the development of new models or theories
and new sophisticated numerical codes. 
Since these codes are complex, extensive testing made through code
inter-comparisons and comparisons with laboratory experiments are
very useful.\cite{2002ApJS..143..201C} Therefore, code benchmarking
can be viewed as one motivation for pursuing laboratory high-energy
density experiments.  These experiments can also be considered as
the most relevant approach to check assumptions and to provide hints
and new ideas to address open questions or issues such as radiative
shocks in astrophysics.\cite{1993ARA&A..31..373D}

High-energy density laboratory astrophysics (HEDLA) experiments are
mostly driven on large-scale lasers
\cite{1991PhRvL..66.2738G,1999PhRvL..83.1982F,2000PhRvE..62.8838S,%
1986PRL.bozier,2004.PRL.Bouquet,2001PhPl....8.2439W}
or on Z-pinches.\cite{2002PhPl....9.2186B,2002ApJ...564..113L}
In a first kind of experiment, one measures microscopic quantities
required for the determination of the equation of state and the
opacities of hot and dense matter found in giant planet interiors or
in stellar atmospheres.
In a second kind of experiment, one seeks to reproduce astrophysical
phenomena in the laboratory. The size of the experimental targets is
of the order of a millimeter, implying typical durations of
a few nanoseconds, whereas the scale of astrophysical phenomena are
between 15 and 25 orders of magnitude larger. However, scaling laws
help to bridge laboratory experiments and astronomical
phenomena.\cite{1999ApJ...518..821R,2001PhPl..8.1804R} In order to
be relevant for benchmarking, the appropriate energy requirements
need to be satisfied and, more importantly, a set of accurate time
and space-resolved diagnostics must be provided. Many
experiments have been devoted to the study of typical radiative
hydrodynamical situations, such as {\em (i)} radiative blast waves
in the context of supernovae
remnants,\cite{2000ApJ...538..645K,2002PhRvL..89.125002,2001PhRvL..87h5004E}
{\em (ii)} radiative precursor shocks
waves,\cite{2004.PRL.Bouquet,1986PRL.bozier,2002PhRvL..89p5003K,Leygnac.PhD.2004}
with applications to the studies of stellar jets, pulsating
stars,\cite{1998A&A...333..687F,2000A&A...354..349F} and accretion
shock during star formation,\cite{2002sf2a.confE.188C} and {\em
(iii)} radiatively collapsing jets relevant for protostellar
outflows.\cite{2000PhRvE..62.8838S,2002ApJ...564..113L}
Reviews can be found in
Refs.~\onlinecite{1999Sci...284.1488R,2001PThPS.143..202T,2005PPCF...47A.191R}.

In this paper, we shall concentrate on the modeling of
radiative-precursor shock wave experiments performed with high-power
lasers. By radiative-precursor shock waves, we mean
highly hypersonic shocks that are driven continuously by a piston.
Because of the high shock velocity, the shocked medium is ionized
and emits radiation, which in turn ionizes and heats the cold
unshocked gas and leads to the apparition of a radiative precursor.
The structure of the shock depends on the shock
velocity.\cite{2000ApJS..127..245B} For a given shock velocity, the
temperature of the shocked matter and its mass density are solutions
of generalized Rankine-Hugoniot equations.\cite{2004.EPJ...michaut}
For weakly hypersonic shocks, the mass density varies continuously
throughout the shock. At higher shock velocities, it becomes
discontinuous and then again continuous when the velocity increases
because of the contribution of the radiation pressure and energy
density.\cite{2000ApJS..127..245B,Alter2003.IFSA..954} In the
so-called supercritical regime, the mass density varies
discontinuously, but the temperature remains constant through the
shock~\cite{2002ZeldoRaizer} except in the shock front where the
temperature spikes. The extent of this region is of the order of the
photon mean free path. 

The strong coupling between radiation and hydrodynamics in these
shocks is difficult to model because of the strong gradients and
different length scales involved for hydrodynamics and radiation
transport. Additionally, an accurate set of opacities and equations
of state for a wide range of  plasma conditions is needed.
Departures from local thermodynamical equilibrium (LTE) are also
expected to be important, especially in the shock front, requiring
detailed calculations of the monochromatic radiation intensities. A
complete and detailed study of the shock structure can therefore be
achieved currently only in a restricted, 1D geometry.  However, the
finite radial size $R$ (in the direction perpendicular to the shock
propagation) of actual shocks in the cosmos and in laboratory
experiments may introduce departures from the ideal 1D behavior. The
purpose of this paper is to examine the importance of these 3D
geometrical effects.

In a 1D description, the radial extension $R$ of the region in the
shock that emits radiation is considered to be infinite.
The amount of radiation
absorbed at one point of the configuration (in the precursor for
example) is therefore overestimated by comparison with a description
where $R$ is finite.
This geometrical effect results in a ``radiative energy loss''
because a fraction of the energy is radiated
radially and does not heat the radiative
precursor ahead of the shock. This effect will be dominant when the
photon mean free path becomes large compared to $R$ and may have a
strong impact on the development of the radiative precursor and on
the shape of the shock.

Deficiencies in the current modeling of radiative shock experiments
provide the motivation for a detailed investigation of the
importance of these lateral radiative losses. For instance, Bouquet
\textit{et~al.}~\cite{2004.PRL.Bouquet} have recently reported on
supercritical radiative shocks created with a high-power laser and
the associated modeling work. While hydrodynamical simulations
reproduce the main features of the experiment, questions remain
regarding the precise understanding of the formation and of the
structure of the radiative precursor.

In this paper, we report a study of the importance of radiative
losses on the structure of shocks produced in laboratory
experiments, aiming at understanding the deficiencies of 1D models.
Section~\ref{sec:experiment} briefly recapitulates the results of
Bouquet \textit{et~al.}~\cite{2004.PRL.Bouquet} and some of these
deficiencies. 
An analytical estimation of the effects of the finite size on the
radiation distribution is given in Sec.~\ref{subsec:Analytical}.
The numerical calculation of the spatial structure of the radiation
field in a stationary case without coupling to the fluid is
described in Sec.~\ref{sec:NumerCalc3D}. 
A full description of the radiative shock structure and of the
radiation field is studied using the 2D Lagrangian radiative
hydrodynamics code FCI,\cite{fci1fci2} with various boundary
conditions for the radiation (Sec.~\ref{sec:Bidim}). We conclude
(Sec.~\ref{sec:conclusion}) that it is essential to account for
multidimensional effects in modeling radiative shocks with finite
radial extension.

\begin{figure}[t] 
\includegraphics[width=0.9\largFigSeb]{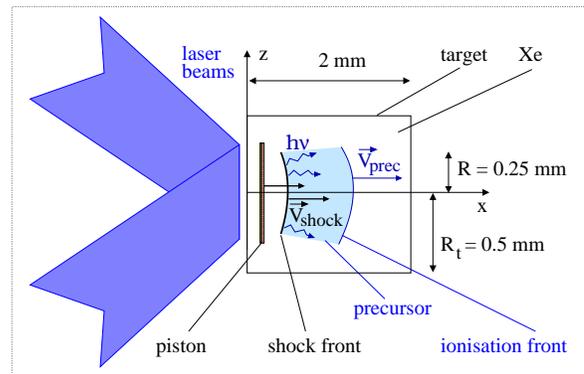}
\caption{\label{fig:target} (Color online) Propagation of the shock
(radius $R$) in the cell (radius $R_\text{t}$) filled with Xe. The
laser beams are shown for illustrative purpose: during the
experiment, the laser pulse is ended when the shock propagates in
the cell.}
\end{figure}

\section{Radiative shock experiments} \label{sec:experiment}

Several shock experiments have been performed with high energy
density lasers to generate supercritical
shocks.\cite{2002..LPB.20..263,2004.PRL.Bouquet,1986PRL.bozier,Leygnac.PhD.2004,2002PhRvL..89p5003K}
In such experiments the laser energy is converted into kinetic
energy of a piston, which drives a shock in a millimetric target of
radius $R_\text{t}$ (see Fig.~\ref{fig:target}) filled with gas
or low-density material.

The extension $R$ (shock radius) of the experimental shock in the
direction perpendicular to its propagation along the $x$ axis (see
Fig.~\ref{fig:target}) is given by the size of the laser spot
focused on the piston. The diameter $2R$ of the focal spot is equal
to $500\;\mu$m [full width at half-maximum (FWHM)] with a
$250\;\mu$m flat center where the intensity is constant. The shock
has the same cylindrical symmetry than the laser focal spot. 

The target is impacted by the laser beams on one of its two ends
consisting in a piston composed of several layers. The first layer is
ablated during the $0.8$ ns duration of the laser pulse. This
ablation produces a shock wave that propagates in the $x$ direction.
The second layer is a radiative shield in titanium that prevents the
radiation of the laser or the radiation emitted during the heating
of the piston by the laser from penetrating the xenon gas.
The shock finally breaks out in the tube filled with xenon gas at
$0.1$ or $0.2$~bar. The shock velocities achieved in the gas are
in the range 50 - 100 km~s$^{-1}$ and the duration of the experiment
is 10 ns.

The spatially averaged temperature of the shock 
is measured\cite{2006PhyPl.13.0702V} to be 10 - 20 eV and the
electron density\cite{2003.IFSA..958} in the precursor is between
$10^{18}$ and $10^{20}$~cm$^{-3}$. The precursor propagates ahead of
the shock front with a velocity that
equals up to twice the velocity of the shock front.

The preparation and the initial analysis of the experiments were
performed using one dimensional (1D) radiation hydrodynamics codes
(FCI,\cite{fci1fci2} MULTI \cite{1986unpu.ramis,1988CPC..ramis}).
While the experimental shocks deviate from 1D geometry, these codes
give an overall good agreement with the experiments (shock velocity,
temperature of the shocked material). However, they are only
moderately successful regarding the radiative precusor. They fail to
reproduce the shape of the electron density profile in the precursor
on the whole interval of values accessible to measurement (from
$10^{18}$ to $10^{20}$ cm$^{-3}$ in
Ref.~\onlinecite{2004.PRL.Bouquet}).
The different codes provide a range of predicted velocities for the
precursor. For example, for a particular shot, the velocity at
$n_{e} = 3\,10^{19}$ cm$^{-3}$ is 300 km~s$^{-1}$ in the 1D version
of FCI and 120
km~s$^{-1}$ in MULTI, the last one being in accordance with the
measured velocity. It was reported by Vinci {\sl et
al.}\cite{2006PhyPl.13.0702V} that the time evolution of the radial
extension of the shock was well reproduced by the 2D version
of FCI, therefore showing that 2D effects have to be considered.

In the work presented here, we estimate the 2D effects and the
consequences of the finite lateral size $R$ of the shock on the
radiation field. When investigating the geometry of the problem, one
needs actually to consider the optical depth, $\tau$, which is the
geometrical distance normalized by the photon mean free path. The
opacity of the medium is therefore an essential parameter of the
problem. Considering the overall agreement of the 1D codes with
the experimental results, we may assume that the current opacities
are a reasonable approximation.
As a consequence, we will not examine that specific point further.

%
%
\section{\label{sec:TR} Radiation field emitted by a shock}
\subsection{Analytical estimation of the multidimensional effects}
\label{subsec:Analytical}

\begin{figure}[t] \centering
\includegraphics[width=0.8\largFigSeb]{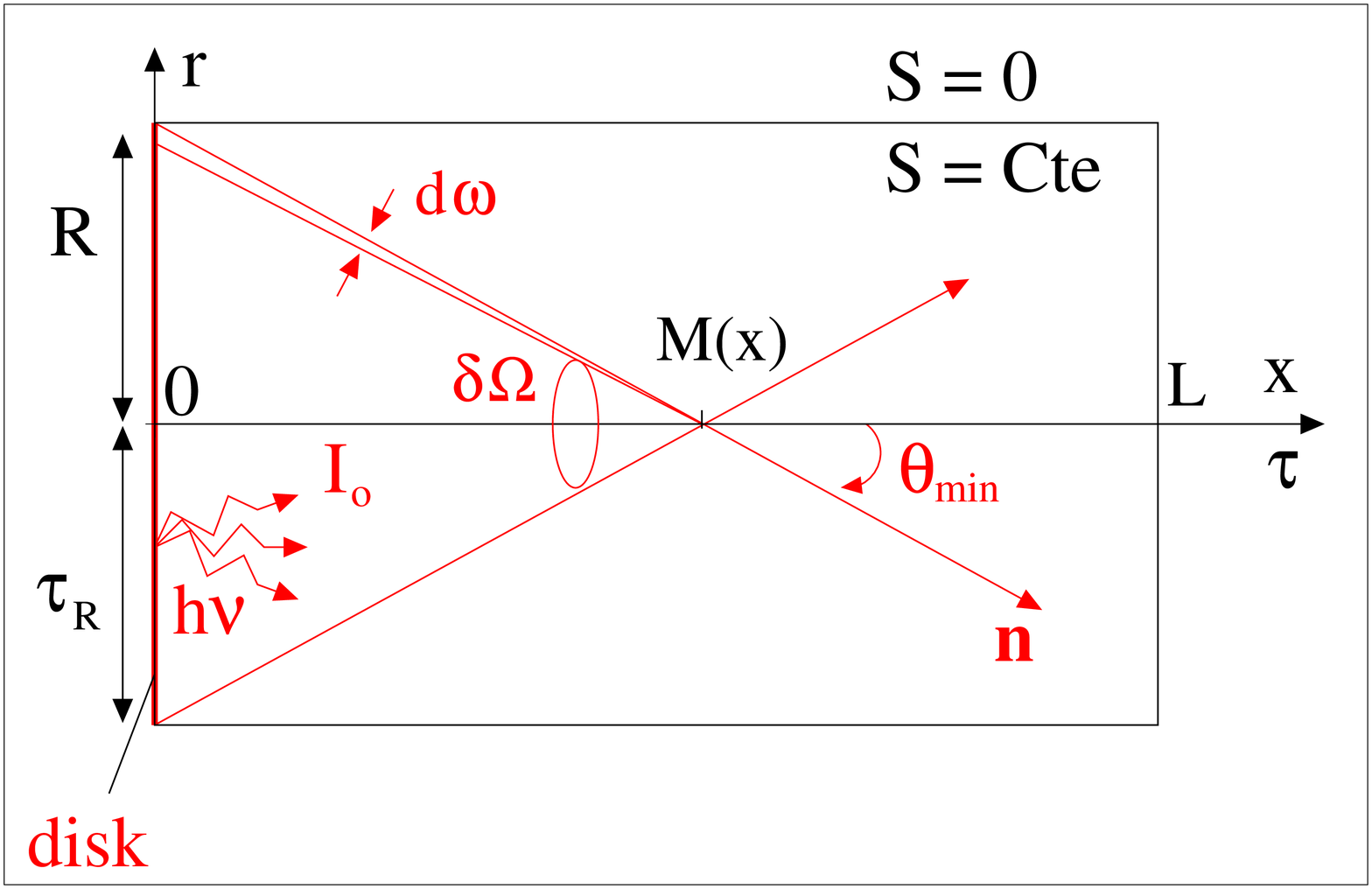}%
\caption{\label{SL:radloss:1geo} (Color online) Geometry of the
configuration for the analytical model. The disk is located at
$x=0$. The limit angle $\theta_\text{min} =
\cos^{-1}{\mu_\text{min}}$ is the angle under which the disk is seen
from the point M(x).}
\end{figure}

We start with an analytical study of the radiation field
generated by a disk in a grey and homogeneous cylinder with radius $R$
and length $L$ (Fig.~\ref{SL:radloss:1geo}).
We furthermore consider that the opacity $\chi$ (cm$^{-1}$, the
inverse of the photon mean free path) is uniform (independent of
space) and we neglect scattering for sake of simplicity.
The optical depth in the $x$ direction $\tau(x) = \int_0^x{\chi
dx}$, is the position along the $x$ axis normalized by the opacity.
The cylinder length $L$ is such that $\chi L = 100$. 
The disk radiates an isotropic specific intensity $I_o$
(J~m$^{-2}$~s$^{-1}$~sr$^{-1}$) and is characterized by its radius
$R$ or by its lateral optical depth $\tau_R = \chi R$.
The local source function $S$ (J~m$^{-2}$~s$^{-1}$~sr$^{-1}$) is
uniform in the cylinder and much lower than the specific intensity
$I_o$. It is zero outside of the cylinder.

In this approach, we would like to calculate the fraction of
radiation energy emitted by the disk and received at any point M(x)
of the cylinder axis $x$. In addition, it is aimed to examine the
way this amount of energy varies by changing the disk radius. A
brief presentation of this calculation has been presented in
Ref.~\onlinecite{2006JPhy4.133..453L} and an extensive version can
be found in Ref.~\onlinecite{Leygnac.PhD.2004}.
%
%
%
We will use the three Eddington variables: the mean intensity $J$,
the flux $\bf H$, and the tensor $\tens{K}$ analogous to the pressure
tensor. They are, respectively, related to the radiation energy
density $E$, the radiative flux ${\bf F}$, and the radiative pressure
tensor $\tens{P}$ through the relations (see
Ref.~\onlinecite{1978stat.book.....M}):
\begin{equation}    
  \label{eq:defJHK} \left(
  \begin{array}{c}
          J  \\ 
     {\bf H} \\ 
    \tens{K}
  \end{array} \right) ({\bf r})
= \frac{1}{4\pi} \left( 
 \begin{array}{c} 
          c E  \\
       {\bf F} \\ 
    c \tens{P} 
  \end{array} \right) ({\bf r})
= \frac{1}{4\pi}  \oint \left( 
  \begin{array}{c} 1 \\ 
    {\bf n} \\ 
    {\bf nn} 
  \end{array} \right)
I({\bf r}, {\bf n}) \, d\omega ,
\end{equation}
where $c$ is the speed of light in vacuum.
In these expressions, the integration is performed over the solid
angle $\omega$, and $I({\bf r}, {\bf n})$ is the specific intensity
at position M({\bf r}) in the direction given by the unit direction
vector {\bf n}. Equations \eqref{eq:defJHK} are frequency-dependent,
but since we assume that the opacity $\chi$ is grey, we will
consider that the radiative quantities appearing in
Eqs.~\eqref{eq:defJHK} are integrated over frequencies.
We will refer to $J$, $\bf H$ and $\tens{K}$ as the radiative
moments since they correspond respectively to the zeroth-, first-
and second-order moments of the intensity over angles.

The formal solution\cite{1978stat.book.....M} of the radiation
transfer equation leads us to decompose the radiation field at
position $M({\bf r})$  in two components: the contribution $J_{o}$,
${\bf H}_{o}$ and $\tens{K}_{o}$ coming from the disk, and the
emission $J_\text{S}$, ${\bf H}_\text{S}$ and $\tens{K}_\text{S}$ by
the uniform medium. The total radiation field is
\begin{equation}   
 \left(
  \begin{array}{c}
          J  \\
     {\bf H} \\
    \tens{K}
  \end{array} \right) ({\bf r})
    =
    \left(
       \begin{array}{c}
              J_{o}  \\
        {\bf H}_{o}  \\
       \tens{K}_{o}
       \end{array}
    \right)
    +
    \left(
       \begin{array}{c}
            J_\text{S}        \\
      {\bf H}_{\text{S}}      \\
     \tens{K}_{\text{S}}
       \end{array}
    \right).
\end{equation} 

In this simple model, it is straightforward to calculate the
radiation field existing at point $M(\tau)$  on the $x$ axis
of the cylinder ($r=0$). 
The contributions of the source function, $J_S$, $H_{x\, S}$ and
$K_{xx\, S}$, are calculated by summing the emission coming from
all the points of the cylinder.
The contribution of the disk is given by
\begin{equation}
\label{eq:JHK_disk}
    \left(
       \begin{array}{c}
       J_o        \\
       H_{o\, x}  \\
       K_{o\, xx}
       \end{array}
    \right)(\tau,r=0)
    = \frac{I_o}{2}  \int^1_{\mu_\text{min}}
    \left(
       \begin{array}{c}
       1 \\
       \mu\\
       \mu^2
       \end{array}
    \right) e^{-\tau/\mu} d\mu ,
\end{equation} \\
where we have considered only the $x$ component of the flux and the
$xx$ component of the tensor $\tens{K}_o$. The quantity $\mu$ is
defined by $\mu = \cos{\theta}$, where $\theta$ is the angle between
the direction $\bf n$ and the $x$ axis.
The integrals given by Eqs.~\eqref{eq:JHK_disk} are exponential
integrals in the case $\mu_\text{min} = 0$ corresponding to a disk
with a radius $R$ going to infinity.  For a finite radius of the disk,
the specific intensity is integrated over the solid angle
$\delta\Omega$ under which we see the disk from point~$M(\tau)$
(see Fig.~\ref{SL:radloss:1geo}).
%
%
%
\begin{figure}[t!] \centering
 \includegraphics[width=\largFigSeb]{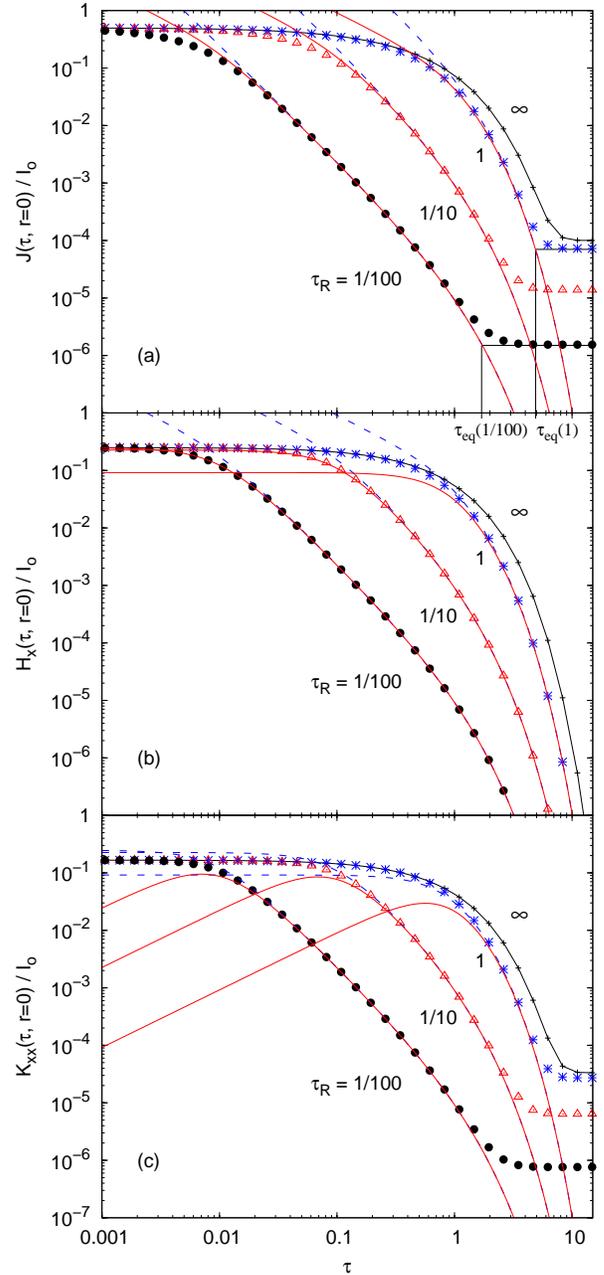}
\caption{\label{SL:radloss:1} (Color online) Radiative moments (a)
$J/I_o$, (b) $H_x/I_o$ and (c) $K_{xx}/I_o$ produced by a disk in a
grey and homogeneous medium with $S=10^{-4}I_o$. The exact solution
is shown for the infinite case (plain line with crosses) and for
$\tau_R=1$, 1/10 and 1/100 (stars, triangles and big dots). Also
shown are the approximate solutions given by
Eqs.~\eqref{eq:JHK_disk_approx} (dashed lines) and
Eqs.~\eqref{eq:JHK_disk_approx} times $\mu_\text{min}$ (plain
curves), which are given by Eqs.~\eqref{eq:JHK_disk_approx2} for $J$
and $H_x$. Positions of $\tau_\text{eq}(\tau_R)$ are shown on frame
(a) for $\tau_R=1$ and 1/100.}
\end{figure}

The total value $J (\tau) = J_o(\tau) + J_S(\tau)$ (and similarly
for $H$ and $K$) is plotted in Fig.~\ref{SL:radloss:1}.
Because the general behavior of $J$, $H_x$ and $K_{xx}$ is similar,
we can restrict our description mostly to the mean intensity $J$.
The decrease of the radiative moments with $\tau_R$ enables us to
understand the effect of the finite radial extension $R$ of the
medium on the structure of the radiation field.

Far from the disk, the radiation emitted by the disk has been
considerably absorbed, and the moments depend only on the local
source function $S$. In the calculations shown in
Fig.~\ref{SL:radloss:1}, the ratio $S/I_o = 10^{-4}$. This value
corresponds to a system in which the radiation is due to Planck
emission and the temperature of the medium is 10 times lower than
the disk temperature.
The flux $H_x(\tau)$ goes to zero when $\tau$ is greater than 1.
An order of magnitude of $J_S$ can be estimated by assuming the
medium is a sphere of radius $R$,
\begin{equation} \label{eq:Estim_JS}
J_S(\tau\gg 1) 
\sim \dfrac{1}{2}\int_{-1}^1{d\mu\, \int_{0}^{\tau_R}{d\tau \, S \, e^{-\tau}}}
\sim S \left( 1 - e^{-\tau_R} \right).
\end{equation}
Though a rough estimate, it is quite correct in the cases considered
here.  For small  $\tau_R$, the expansion of $e^{-\tau_R}$ gives 
$J_S(\tau\gg 1) \sim S \, \tau_R$,
an evaluation quite close to the exact value [see in
Fig.~\ref{SL:radloss:1} the value of $J(\tau>10)\sim J_S\sim S
\tau_R$].

An interesting feature of the decrease of $J$ with $\tau_R$ can be
illustrated with Fig.~\ref{SL:radloss:1}: for $\tau_R$
smaller than one, it takes a distance of a few $\tau_R$ for
$J$ to be absorbed by a factor of 10,
whereas, in the infinite case, where $\tau_R$ is greater than one, the
radiation is absorbed tenfold after a propagation of a few
mean free paths of photons, i.e.  at $\tau \approx 1-2$. The
absorption of the radiation field is therefore determined by the
smallest of the characteristic scales. For the system studied here,
the smallest scale is either the mean free path of photons $1/\chi$
or $R$.

In the case $\tau_R = 1$ for which the two scales are the same,
$J_o$ is not far from the mean intensity $J_{0\, \infty}$ calculated
in the infinite case. At position $\tau = 2$, $J_o$ is only a factor
of four lower than $J_{0\, \infty}$. Therefore, a characteristic
lateral optical depth $\tau_R$ of a few is sufficient to consider
the medium as optically thick and consequently nearly identical to a
medium with an infinite lateral dimension.

The other two moments for the disk radiation $H_{0\, x}$ and $K_{o\,
xx}$ have similar profiles, and the three moments $J_o$, $H_{0\, x}$
and $K_{o\, xx}$ have nearly the same value sufficiently far away
from the disk, where it is possible to have an approximate solution
of Eqs.~\eqref{eq:JHK_disk}.
At a distance from the disk of a few times $\tau_R$, the solid angle
$\delta\Omega$ is small and
$\mu_\text{min} 
\approx 1 -        \theta^2/2 
\approx 1 - (\tau_R/\tau)^2/2$.
The limiting value $\mu_\text{min}$ is close to unity as soon as
$\tau$ is larger than 3 $\tau_R$ or 4 $\tau_R$ and asymptotically
goes to unity.

The limits of the integrals in Eqs.~\eqref{eq:JHK_disk} are
%
\begin{equation}    
\label{eq:JHK_disk_approx}
    \left(
       \begin{array}{c}
       J_o \\
       H_{o\, x} \\
       K_{o\, xx}
       \end{array}
    \right) (\tau,r=0)
    \underset{\mu_\text{min} \rightarrow 1}{\longrightarrow} 
    \dfrac{I_0}{4} \left( \dfrac{\tau_R}{\tau} \right)^2\,
                          e^{-\tau/\mu_\text{min}}
    \left(
       \begin{array}{c}
       1 \\
       \mu_\text{min}\\
       \mu_\text{min}^2
       \end{array}\right).
\end{equation}

This asymptotic expression (represented in Fig.~\ref{SL:radloss:1})
is useful to approximate the exact Eqs.~\eqref{eq:JHK_disk} over a
range $[\tau_0, \tau_\text{eq}]$ and to describe more simply the
absorption of the radiative moments. Let us first estimate the lower
value $\tau_0$. At position $\tau = 5\tau_R$, the error on $J_o$
defined as the relative difference $(J_o^\text{approx} -
J_o^\text{exact})/J_o^\text{exact}$ is, respectively, 6\%, 0.6\% and
0.06\% for, respectively, $\tau_R=1$, $0.1$ and $0.01$. We see on
Fig.~\ref{SL:radloss:1} that $\tau_0$ is of the order of a few
$\tau_R$.
Therefore, the approximation stands only far enough from the disk,
where the mean intensity has been greatly reduced by the absorption.
For example, $J_o(\tau = 5\tau_R)/J_o(0) \approx 1/100$ and $1/50$
for respectively $\tau_R=0.1$ and $0.01$.

Let us define $\tau_\text{eq}(\tau_R,I_o/S)$ as the optical depth
for which the disk radiation is equal to the local radiation. This
is a straightforward way to separate the medium in a region
$\tau<\tau_\text{eq}$ where the radiation field coming from the disk
dominates and the region $\tau>\tau_\text{eq}$  where the disk does
not have a direct influence on the radiation field. The transition
between the two regions is sharp because the radiation field emitted
by the disk decreases exponentially around $\tau_\text{eq}$. This
$\tau_\text{eq}$ can be easily calculated numerically. Its
approximate position is shown Fig.~\ref{SL:radloss:1}.a for
$\tau_R=1$ and 1/100.
We find that $\tau_\text{eq}$ decreases slower than $\tau_R$ since
the local radiation $J_S$ decreases like $\tau_R$ when $\tau_R$ is
small [see Eq.~\eqref{eq:Estim_JS}], whereas the disk radiation
decreases like $(\tau_R/\tau)^2 \exp(-\tau)$.
We see in Fig.~\ref{SL:radloss:1} that $\tau_\text{eq} \sim 5$ for
$\tau_R = 1$ and $\tau_\text{eq} \sim 2$ for $\tau_R = 1/100$.
Therefore, $\tau_\text{eq}$ decreases by less than a factor 5 when
$\tau_R$ decreases by a factor 100.
Then, in this particular cylindrical configuration, the range
$[\tau_0, \tau_\text{eq}]$ over which the approximation
\eqref{eq:JHK_disk_approx} is good increases as $\tau_R$ decreases
since $\tau_0$ decreases proportionally to $\tau_R$ and
$\tau_\text{eq}$ does not vary a lot.

Equations \eqref{eq:JHK_disk_approx} show that the radiative moments
on the $x$ axis are absorbed by a factor $e^{-\tau/\mu_\text{min}}$
and vary like $(\tau_R/\tau)^2$ which, for a uniform opacity, is the
solid angle $(R/x)^2$ under which the disk is seen from
position~$\tau$.

Furthermore, we can calculate from Eqs.~\eqref{eq:JHK_disk_approx}
the Eddington factor $f = K_{o\, xx}/J_o \approx \mu_\text{min}^2$
and the anisotropy factor $H_{o\, x} / J_o \approx \mu_\text{min}$.
We conclude that the radiation field has a high degree of anisotropy
at distances where the approximation is acceptable.

The approximation \eqref{eq:JHK_disk_approx} can be improved for
$J_o$ and $H_{o\, x}$ (but not for $K_{o\, xx}$) if they are
multiplied by $\mu_\text{min}$,
\begin{equation}    
\label{eq:JHK_disk_approx2}%
    \left(%
       \begin{array}{c}%
       J_o \\
       H_{o\, x}
       \end{array}%
    \right) (\tau,r=0)
\approx
    \dfrac{I_0}{4} \left( \dfrac{\tau_R}{\tau} \right)^2\,
                          e^{-\tau/\mu_\text{min}}
    \left(
       \begin{array}{c}
       \mu_\text{min}\\
       \mu_\text{min}^2
       \end{array}\right).
\end{equation}
The error for $J_o$ is then decreased by a factor of 5 to 10, and
this approximation is valid for a larger range than approximation
\eqref{eq:JHK_disk_approx} (see Fig.~\ref{SL:radloss:1}), namely for
values of $\tau$ greater than approximately $\tau_R / 2$.
For $H_{o\, x}$, the improvement is very good since the error is
then lower than $\tau_R$ for all values of $\tau$. It is therefore
very useful for $\tau_R < 0.1$, since the accuracy is then better
than $10 \%$.
With this new approximation, the Eddington factor becomes $f =
K_{o\, xx}/J_o \approx \mu_\text{min}$ and, when compared with the
exact value, proves to be more accurate than $\mu_\text{min}^2$.

\subsection{Numerical calculation of the radiation field generated
by a planar shock}
\label{sec:NumerCalc3D}
\subsubsection{Multidimensional radiative transfer with the short
characteristics method}

The above analytical calculation is limited to the values of the
radiation field on the $x$ axis in a uniform medium with a grey
opacity. We now perform a multidimensional numerical
computation of the radiation field generated at a given time by a
planar shock. We vary the lateral size of the shock and study how
the radiation field is modified while all the other quantities are
kept invariant.
The coupling of the radiation and the fluid will be studied in
Sec.~\ref{sec:Bidim}. For the moment, we are only interested in the
shape and the value of the radiation field in a nongrey
description.

\begin{figure}[t!]
%
%
  \includegraphics[width=\largFigSeb]{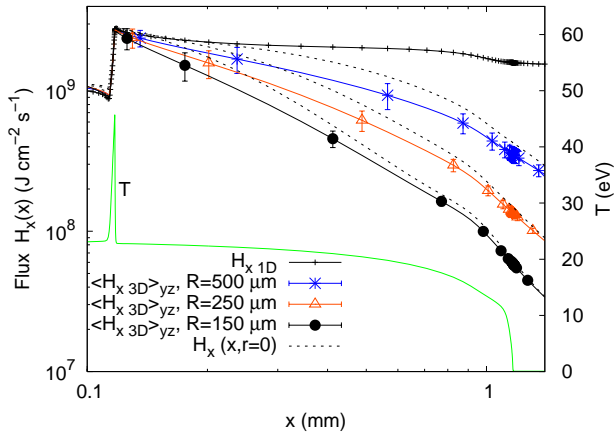}%
  \caption{\label{SL:radloss:2} (Color online) 3D and 1D numerical
computation of the flux $H_x$ (from the model presented in
Sec.~\ref{sec:Compar1D3D}) for several values of $R$. The lines with
symbols are the mean values, the dashed lines are the values at
position $r=0$, and the error bars are the mean deviations. The
temperature $T(x)$ of the shock is plotted.}
\end{figure}

The structure of the shock (temperature and density) along $x$ is
given by a one-dimensional radiation hydrodynamics code developed by
J.~P.~Chi\`eze and that has also been used in
Ref.~\onlinecite{2004.PRL.Bouquet}.
It is chosen to be as close as possible to the laboratory shocks we
have observed\cite{2004.PRL.Bouquet} in xenon (velocity $\sim
60$~km~s$^{-1}$, $P = 0.2$~bar). The temperature is shown in
Fig.~\ref{SL:radloss:2}.
This 1D shock structure is put in a three-dimensional (3D) grid by
assuming a plane parallel shock structure.

In the 3D radiative transfer calculation, the medium is assumed to
be in LTE. The source function is therefore assigned to be the
Planck function. The opacity is approximated by a screened
hydrogenic model, and calculated at $n_\nu = 200$ frequencies.

The radiative transfer code calculates the specific intensity
$I({\bf r}, {\bf n})$ at each grid point of a 3D Cartesian grid
(number of grid points: $n_x=200, n_y=40, n_z=40$) with the short
characteristics method.\cite{1988JQSRT..39...67K}
We use the cartesian coordinates because a solution of the radiation
transport in these coordinates is easier to express and faster to
compute than in the equivalent ``natural'' cylindrical system (see
also Ref.~\onlinecite{2002ApJ...568.1066V}).
The smallest radial size of the shock for which the calculation has
been performed is $\Delta Y =\Delta Z = 2R = 300\ \mu$m, i.e., the
size of the experimental shock. For this value of $R$, and for a
cell length $\Delta x = 2$~mm, the number of points for the angles
is $n_\mu = 200$, $n_\varphi = 64$ where the range of $\mu$ is
$[-1,1]$ and the range of $\varphi$ is $[0,2\pi]$. 
Solving the radiation transport in 3D is much more demanding than in
1D not only because of the larger number of grid points ($n_x$,
$n_y$ and $n_z$) but also because of the required angular sampling
($n_\varphi$ and $n_\mu$) which must be more refined as $R$
decreases. Generally, this implies that some trade-off such as
limiting the number of frequencies describing the opacity is
necessary. In the 1D case, $10^5$ to $10^6$ frequencies can be
easily managed, whereas this number is drastically reduced from 10 to
$10^3$ in 3D calculations (here, $n_\nu = 200$).
In this calculation, the total number of mesh points is
$n_x\, n_y\, n_z\, n_\mu\, n_\varphi\, n_\nu \sim 10^{12}$.

\subsubsection{Curvature of the radiation field}
\begin{figure}
\centering
\includegraphics[width=\largFigSeb]{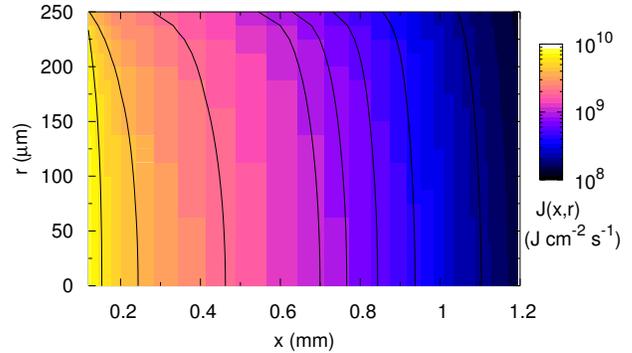}
\caption{\label{fig:SC3D_Jcol_Xe_JXEc1n20m2p64} (Color online) 2D
plot of the mean intensity $J(x,r)$ (J~cm$^{-2}$~s$^{-1}$) for a
planar shock of width $R=250\, \mu$m. Isocontours are drawn for
$J =    2.10^8, 4.10^8,      6.10^8, 8.10^8,
  10^9, 2.10^9, 4.10^9$ and $6.10^9$ J~cm$^{-2}$~s$^{-1}$.
}
\end{figure}

The curvature of the radiation field is evidenced in 
Fig.~\ref{fig:SC3D_Jcol_Xe_JXEc1n20m2p64}. This curvature is due
only to the finite lateral size of the medium since the source
function is kept planar, that is $S(x,r) = B(T(x))$, with $B(T)$ the
Planck function.
The profiles of $H_{x}$ and $K_{xx}$ are similar to the profile of
$J$. At position $x = 500 \ \mu m$ in the radiative precursor, at
mid distance between the shock front and the ionization front, the
value of the mean intensity on the boundary $J(x,R)$ is roughly a
factor of two lower than $J(x,r=0)$, the value on the axis. The
medium is then in an intermediate state between optically thick and
optically thin. At places $x\lesssim 200 \,\mu$m  where the medium
is optically thick, for example in the shock front, $J$ is nearly
constant, which makes it identical to the 1D case. The medium
becomes optically thinner as we go farther away from the shock
front. The radiation field deviates from the planar geometry
and less radiation is available to ionize the medium on the outer
border of the ionization front ($r\sim R$) than on the axis.  This
leads the ionization front to also depart from a plane. This result
is confirmed by the subsequent calculations of Sec.~\ref{sec:Bidim}.

%
%
%

\subsubsection{\label{sec:Compar1D3D} Comparison between 1D and 3D} 
A comparison between the radiation field in 3D and in 1D can be done
by computing its value averaged on the surface $\Delta Y \Delta Z$
at each position $x$:
\begin{equation*}
<\J(x) >_{yz}
= \dfrac{
\int_{-\Delta \text{\tiny Y/2}}^{\Delta  \text{\tiny Y/2}}{dy 
\int_{-\Delta \text{\tiny Z/2}}^{\Delta  \text{\tiny Z/2}}{dz \ \J(x,y,z)}}}
      {\Delta \text{Y} \, \Delta \text{Z}}
\end{equation*}
where $\J$ is either the mean intensity $J$, a component of the flux
$H_{i}$ or a component of the tensor $\tens{K}_{ij}$, where
$i,j=x,y,z$.

The flux $H_x$ calculated with the 3D model
for $R=150$, 250 and 500 $\mu$m and
in 1D for $R = \infty$ is shown Fig.~\ref{SL:radloss:2}. The error
bars are standard deviations on the lateral surface 
and the dashed lines give the value at $r=0$ for each value of $R$.
The magnitude of the radiation field decreases with the lateral size
of the shock, as was shown by our analytical approach. Although the
radiation field is nearly constant in the 1D infinite calculation,
it decreases by one order of magnitude between the shock front (the
spike of temperature) and the ionization front (where the
temperature goes to zero) for $R=500 \ \mu$m and by two orders of
magnitude for $R=150 \ \mu$m. We should note, however, that these
calculations most likely overestimate the radiation field since the
temperature profile is kept the same for all the values of $R$. But
the decrease of the radiation field should result in a cooler and
shorter precursor, which will therefore emit even less radiation than
the 1D infinite-$R$ profile. We are going to see in
Sec.~\ref{sec:Bidim} that when $R$ decreases, the temperature in the
precursor decreases too.


%
%
\section{\label{sec:Bidim}Bidimensional time-dependant numerical
simulations}

In the general case, the determination of the radiation field [for
example by solving Eqs.~\eqref{eq:JHK_disk}] is difficult because
the value of the optical depth $\tau({\bf r})$, which depends on the
opacity $\chi({\bf r})$, must be known. But the opacity depends on
the local density, temperature and the ionization and excitation
state of the medium, which can be modified by the radiation field.

In the previous sections, we studied the consequences on the
radiation field of the variation of the lateral size of the medium,
without calculating the effect on the structure of the medium. We
now focus on the consequences of the bidimensional shape of the
shock on the radiative precursor structure.
We first compare 1D and 2D numerical simulations with various
boundary conditions with respect to radiation, and investigate the
transition between these two kinds of calculations, stressing the
role of the radial (transverse) radiative flux. In a second part, we
study the 2D structure of the radiation field.

\subsection{\label{subsec:trans1D2D} Transition from 1D to 2D}
\begin{figure}
\includegraphics[width=0.85\largFigSeb]{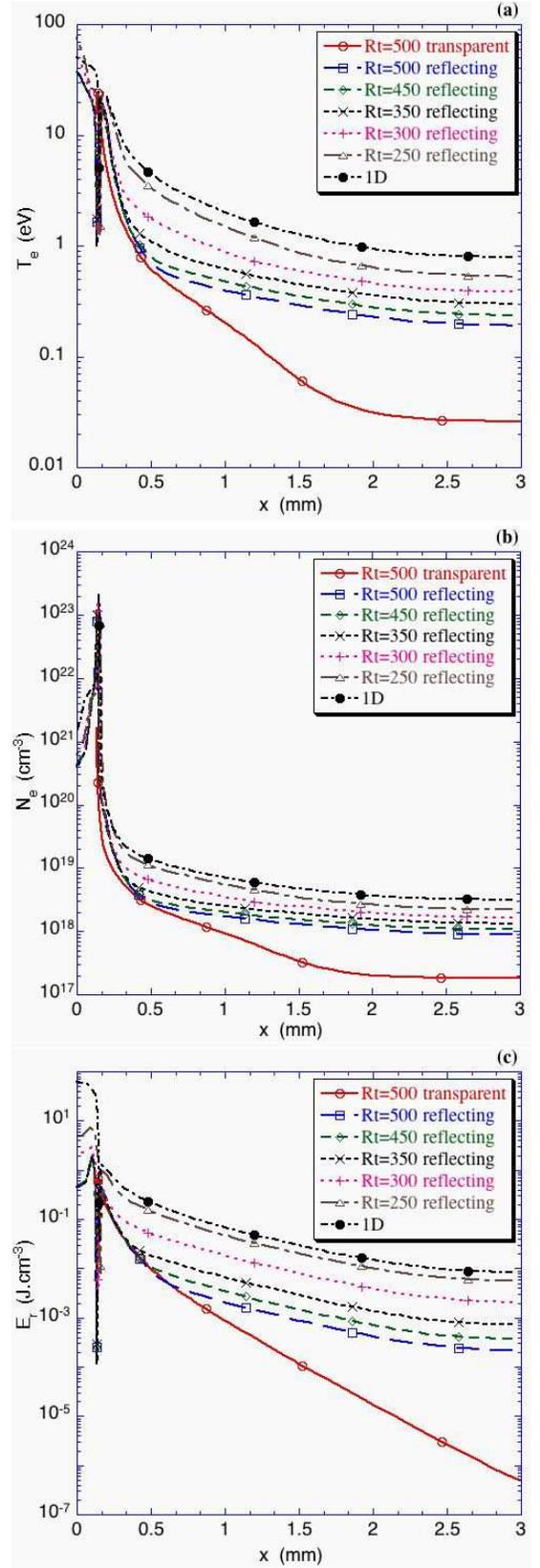}
\caption{\label{fig:TeNeErad1D2Dprofile} (Color online) Transition
from the FCI-1D reference simulation to the FCI-2D reference case,
using decreasing target radius $R_t$ (in $\mu$m) and reflecting wall
cells. Axis profiles of (a) electron temperature (eV), (b) electron
density (cm$^{-3}$) and (c) radiation energy density (J.cm$^{-3}$)
are plotted at t = 3~ns.}
\end{figure}

We perform time-dependant 2D radiation hydrodynamics simulations in
cylindrical geometry with the code FCI.\cite{fci1fci2} This code,
devoted to the modeling of inertial confinement fusion, is therefore
well suited to simulate the radiative shock experiments described in
Sec.~\ref{sec:experiment} from the laser pulse to the shock
propagation in the Xe gas.
The radial and temporal laser profiles can be well approximated by
constant functions, with Gaussian wings and $500\ \mu$m FWHM. The
value of the laser flux ($3.10^{13}$ W/cm$^{2}$ at peak power in
time) is chosen so that the velocity of the produced shock ($55$
km/s) matches the experiment shock velocity.

We define a reference 2D simulation similar to a typical shot
performed in the 2002
campaign.\cite{2002..LPB.20..263,2004.PRL.Bouquet} The target
diameter is 1~mm and the walls of the cell, which were made of
quartz, are transparent to radiation.

The transition between 2D and 1D has been investigated by performing
2D simulations with walls reflecting radiation and decreasing the
target radius $R_\text{t}$. However, the laser profile and the
shock radius $R$ are the same in all simulations. Reflective walls
prevent lateral radiative losses at the boundary
$r=R_\text{t}$.
A 2D simulation with reflective walls and $R_\text{t}$ equal to
the laser spot radius $R$ is nearly equivalent to a 1D simulation.
On the contrary, a calculation with $R_\text{t}$ much larger
than $R$ clearly exhibits the 2D aspects of the problem. Increasing
$R_\text{t}$ in these simulations is then comparable, although
not strictly equivalent, to decreasing the shock radius in the model
of Sec.~\ref{sec:TR}. As stated earlier, the geometrical effect of
decreasing $R$ can be described as a radiative loss. 

This procedure emphasizes the importance of radiative losses both in
simulations of radiative shocks and in related experiments by
varying only one geometrical parameter.

\begin{figure}
   \includegraphics[width=0.9\largFigSeb]{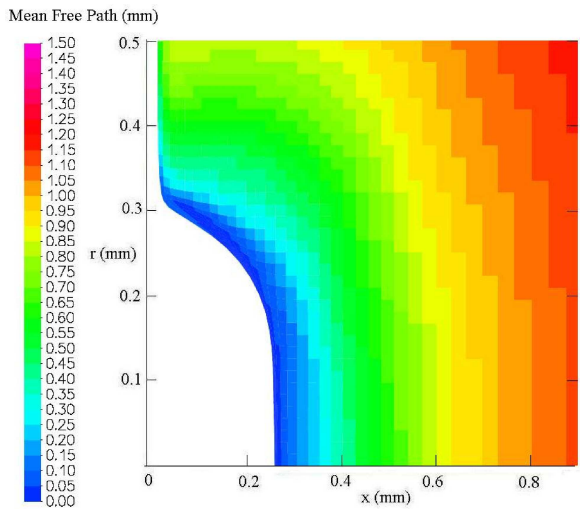} 
    \caption{\label{lpm2D} (Color online) 2D plot of Rosseland mean
free path in mm, in the reference 2D case, at t = 5~ns.}
\end{figure}
In Fig.~\ref{fig:TeNeErad1D2Dprofile}.a, we compare the axial
temperature profiles at t=3~ns in the reference 2D simulation (lower
curve) and the reference 1D simulation (upper curve) computed  with
the 1D version of the code.\cite{fci1fci2} Intermediate plots give
electron temperature profiles for 2D simulations using reflective
walls and decreasing target radius $R_\text{t}=$ 500, 450, 350,
300, and 250 $\mu$m.
Reflecting boundaries suppresse radiation losses and give higher
radiative flux and energy in the precursor
(Fig.~\ref{fig:TeNeErad1D2Dprofile}.c), which then produce higher
temperatures (Fig.~\ref{fig:TeNeErad1D2Dprofile}.a) and electron
densities (Fig.~\ref{fig:TeNeErad1D2Dprofile}.b).
As the radius of the reflecting wall simulations decreases, the
resulting temperature profile gets closer and closer to the
reference 1D profile. 
With reflecting walls and a $350\;\mu$m target radius, the axial 2D
electron density in the precursor is about half that of the 1D
value. With reflecting walls and a  $250\;\mu$m radius, the 2D
simulation yields an axial profile of the electron density quite
similar to the 1D simulation, as expected. Radial flux losses thus
seem to be the main source of differences between 1D and 2D
simulation results.

One can also notice that the difference between the reference 2D
profiles and the others increases farther away from the shock front.
This can be explained by the cumulative effect of radial losses on
the radiative flux emitted by the shock front.  

\begin{figure}[t!]
   \includegraphics[width=0.8\largFigSeb]{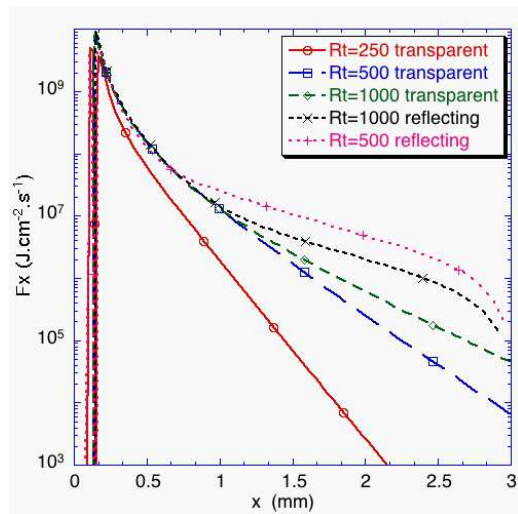} 
    \caption{\label{Fradzaxe2Dprofile} (Color online) Comparison of
axis longitudinal radiative flux (J~cm$^{-2}$~s$^{-1}$) with various
boundary conditions, at t = 3~ns.}
\end{figure}
%
A cartography of the Rosseland photon mean free path in xenon is
shown Fig.~\ref{lpm2D} for the reference 2D simulation at t=5 ns,
when the shock position is $x \approx 250\, \mu$m.  Only the xenon
gas has been represented in the figures.
The region in white color corresponds to the xenon gas, which has
been swept up by the shock initially produced at $x=0$. The radius
of this region is about $300\, \mu$m at the origin $x=0$ whereas the
radius of the laser focal spot is about $250\, \mu$m. We conclude,
therefore, that the radius of the shock, $R$, increases with time
while it is propagating to the right in the xenon still at rest. In
addition, the curvature of the shock front is clearly evidenced.

The values of the Rosseland mean free path $1/\bar\chi$ in the
precursor ($400 \,\mu$m $\lesssim x \lesssim 800 \,\mu$m) range
typically between 200 $\mu$m and 1 mm (or less), which correspond,
respectively, to $\tau_R=\bar\chi R \approx 1.25$ and 0.25, whereas
they strongly decrease in the shocked xenon and the shock front
($250 \,\mu$m $\lesssim x \lesssim 400 \,\mu$m), where $\tau_R\approx
2.5$ for $\bar\chi\approx 0.1$ mm. With these characteristic values
of $\tau_R$, the shock might be considered as nearly 1D in the shock
front and 3D in the precursor.

The effects of boundary conditions on the radiation become less
important as the radius of the target increases, as seen in
Fig.~\ref{Fradzaxe2Dprofile}. The longitudinal flux calculated for
{\bf transparent} walls increases as the radius of the target
increases, and the volume of gas ionized by radiation increases too.
The Rosseland photon mean free path (see Fig.~\ref{lpm2D}) lies
typically between $200\;\mu$m and 1~mm or less in the precursor.
Therefore, most of the radiation emitted by the shock has been
absorbed by the gas before reaching the walls of the cell. The
radiation emitted by all the ionized gas contributes to the
longitudinal flux which increases until the maximum volume of
ionized gas is obtained in the precursor, ahead of the shock.

On the other hand, the longitudinal flux calculated with {\bf
reflective} boundaries decreases as $R_\text{t}$
increases because the radiation that reaches the walls when the
radius is small enough is reflected and then contributes again to
the flux and to the ionization of the gas. The limit situation is
reached when the boundary is sufficiently far away so that most
photons are absorbed in the gas.

\enlargethispage{0.5cm}
\begin{figure}[t!]
   \includegraphics[width=0.9\largFigSeb]{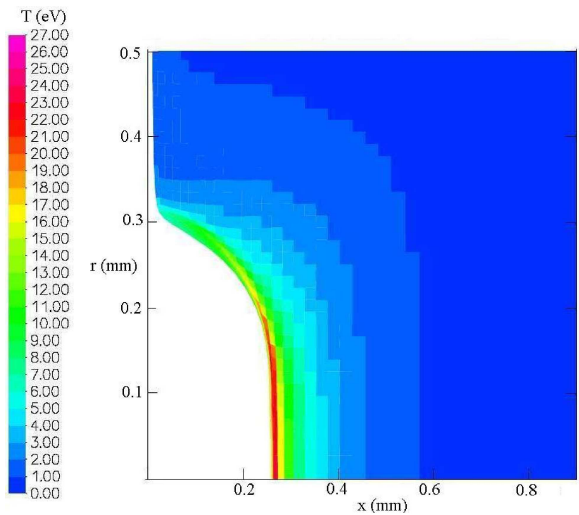} 
    \caption{\label{Te2Dprofile} (Color online) 2D profile of the
electron temperature in the gas in the reference 2D case, at t =
5~ns.}
\end{figure}
%
Another consequence of the smaller amount of radiation emitted in 2D
than in 1D is the smaller precursor velocity.
The precursor velocity, defined as the speed of a chosen isocontour
value of electron density, decreases from 350 km~s$^{-1}$ in the 1D
simulation to 100 km~s$^{-1}$ in the 2D reference case, for $n_{e} =
10^{19}$ cm$^{-3}$.
The 1D simulations thus also overestimate precursor velocities in a
systematic way.

\begin{figure}[t!]
   \includegraphics[width=0.8\largFigSeb]{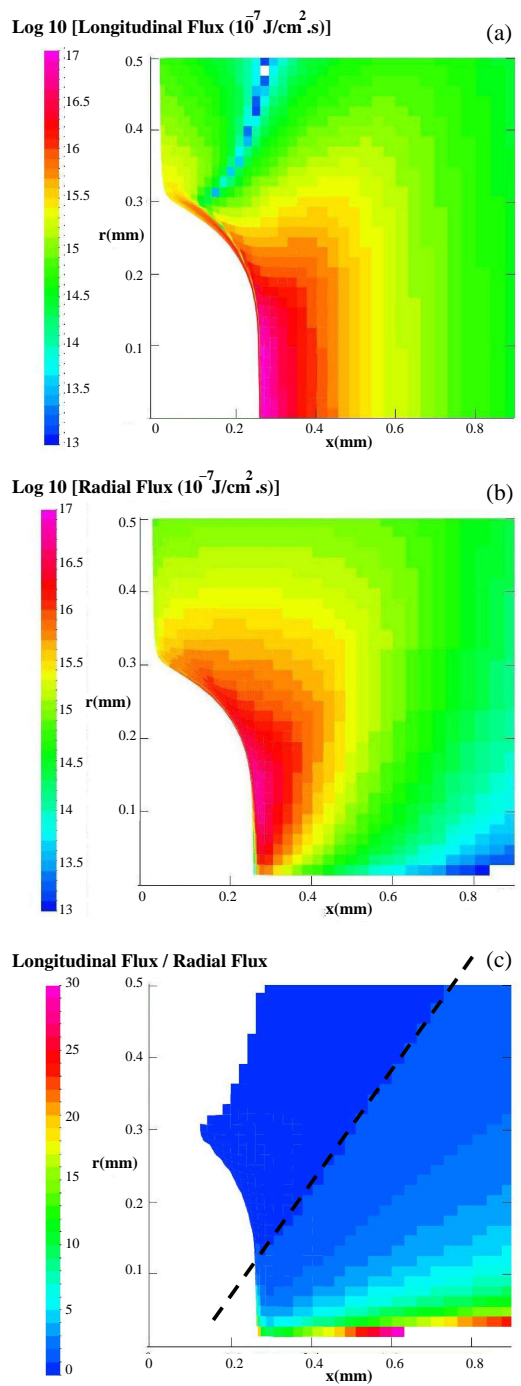} 
    \caption{\label{fig:Fluxes2Dprofile} (Color online) 2D plot of
the (a) longitudinal and (b) radial fluxes (in
J~cm$^{-2}$~s$^{-1}$), and (c) their ratio in the reference 2D case,
at t = 5~ns.}
\end{figure}

\subsection{\label{subsec:2Dradfield} Bidimensional structure of
the radiation field}
Figure \ref{Te2Dprofile} shows that the  electron temperature is much
higher in the shock front than in the precursor: 25~eV versus 5~eV
and less. Assuming that emissivity scales as $T^{4}$ like a black
body, this makes precursor emissivity mostly negligible compared to
the emissivity of the shock front. Since in addition the average
mean free path in the precursor is similar to the cell and precursor
length scales, we can expect the radiation field to be effectively
dominated by the radiation coming from the shock front. This will
not be true anymore far enough from the shock front. The analysis of
Fig.~\ref{Fradzaxe2Dprofile} done in Sec.~\ref{subsec:trans1D2D}
shows that, for $R_\text{t} \geq 500\,\mu$m, the radiation from
the volume of the gas (the precursor) becomes important at distances
from the shock front greater than about 1~mm.

Let us consider the radial and longitudinal components of the
radiative flux, plotted in Fig.~\ref{fig:Fluxes2Dprofile}.a and
Fig.~\ref{fig:Fluxes2Dprofile}.b.
On the $x$ axis, the longitudinal flux is maximum, whereas the
radial flux is zero because of the cylindrical symmetry. The radial
flux is most important near the shock front. On
Fig.~\ref{fig:Fluxes2Dprofile}.c showing the ratio between both
fluxes, we see that the medium can be separated in two regions with
different regimes. On the dashed line, the two components are
roughly equal, while the radial flux dominates above this line and
the longitudinal flux dominates closer to the axis, below this
separation.

The radiation flux field appears, therefore, very anisotropic near
the axis. It is strongly oriented in the shock propagation
direction.
This structure of the radiation field is very similar to that
obtained in the model presented in Sec.~\ref{sec:TR}, which had a
planar shock structure and transparent walls.
This suggests that the 2D radiation at $r < 250\;\mu$m could be well
modelled by a planar uniform source with finite radius $R$
(representing the shock front) and transparent walls.

\section{\label{sec:conclusion} Conclusion}
Geometrical effects on the radiation field are important when the
radial optical depth $\tau_R=\chi R$ is lower than unity.
Analytical arguments explain the decrease of the energy density,
flux, and pressure when the radius $R$ of the emitting surface
diminishes. They show that, when $R$ is lower than the photon mean
free path $1/\chi$, the radiation is absorbed roughly by a factor of
ten at a distance $x$ from the emitting surface of a few $R$. The
dimension of the heated region therefore scales like $R$ in this
simple model.
Approximate solutions of the radiation energy density, flux, and
pressure are found in the asymptotic limit $x \gtrsim R$. Further
studies are needed to estimate their relevance in more complexe
models.
Verifications of the radiative transfer code can also be made by
comparison with the analytical solution.

Numerical calculations of radiative shocks confirm that the radius
of the shock must be taken into account when $\tau_R=\bar\chi R < 1$
even though $\bar\chi$ is a mean opacity (the Rosseland mean in the
present case), which is a very rough representation of the actual
frequency dependence in the radiative transfer calculations.
One-dimensional calculations for such a complex system overestimate
the amount of energy radiated and transfered ahead of the shock.
This results in overestimating the energy deposition in the
precursor. Therefore, all the properties of the radiative precursor
are overestimated: its velocity by a factor 3, the temperature and
electron density by one order of magnitude, and the extension by a
factor that depends on the electron density, but can be up to an
order of magnitude. These values are for the properties of the
precursor far enough from the shock front, say at more than a photon
mean free path.
Moreover, the shape of the radiation field in the precursor and the
structure of the temperature and electron density in the precursor
depart from a plane parallel geometry. Our work, therefore,
emphasizes the need of considering 3D or 2D-cylindrical numerical
simulations for modeling radiative shocks in the laboratory and in
cosmic settings.

\bibliography{bibArt}

\end{document}